\documentstyle[prl,aps,psfig]{revtex}
\begin{document}
\title{Phase-coherent tunneling through a mesoscopic superconductor coupled to
superconducting and normal metal electrodes. }
\author{A.V.Zaitsev $^{\ast}$, A.F. Volkov$^{\dagger}$,
S.W.D.Bailey$^{\dagger}$, and C.J.Lambert$^{\dagger}$}
\address{${\ast}$ Institute of Radioengineering and Electronics of the
Russian\\
Academy of Sciencies, Mokhovaya str.11, Moscow 103907, Russia.\\
${\dagger }$ School of Physics and Chemistry, Lancaster University,\\
Lancaster LA1 4YB, UK.}
\maketitle

\begin{abstract}
Phase-coherent diffusive transport through mesoscopic hybrid
superconductor/normal metal tunneling structures is investigated. For a $%
N-s-S$ two-barrier tunneling system with bulk $S$ and $N$ electrodes coupled
by a mesoscopic superconducting constriction $s$, zero-bias conductance and
non-linear I-V curves are calculated under the assumption that the dwell
times of quasiparticles in the $s$ region is shorter than inelastic
relaxation time. It is shown that the low voltage conductance of this system
determined by the Andreev reflection processes may exceed the conductance in
the normal state and its value is very sensitive to the weak pairing
interaction of electrons in the $s$ region. We show that even weak pairing
electron interaction may result in the significant qualitative and
quantitative change of the conductance temperature dependence with respect
to the case of structures with the normal mesoscopic region. We calculate
the I-V curves and show that they depend on the applied voltage in a
non-monotonic way, therefore differential conductance becomes negative with
increasing voltage. Such behavior is due to the voltage dependence of the
order parameter in the constriction and the phase difference $\varphi $
between the $S$ and $s$ superconductors. It is shown that if the tunneling
processes determine the form of the quasiparticle distribution function in
the $s$ superconductor, the phase $\varphi $ is stationary at arbitrary
voltages. For quasiparticle tunneling interferometers in which the
mesoscopic superconductor, $s$, couples the superconductor, $S$, and the
normal metal, $N$, the zero bias conductance, as a function of the phase
difference between the $S$ electrodes is investigated. It is shown that the
amplitude of the conductance oscillations may exceed the conductance of this
structure in the normal state.
\end{abstract}
\subsection{Introduction}
Phase-coherent transport in mesoscopic superconductor/normal metal ($S/N$)
systems has been an active area of research during the last decade \cite{LR}.
The interest in the theoretical investigations was stimulated by
impressive technological advances and by  experimental activity in studying
various properties of small mesoscopic structures 
\cite{r1,r2,r3,r4,r5,r6,r7,r8,r9}.
Interesting phenomena in mesoscopic systems are due to the
importance of both the phase coherence established in the $s$ constriction 
by the proximity effect and significant
departure of quasiparticles from equilibrium. This is completely true for
two-barrier structures $N-s-S$ with barriers at the interfaces between 
the $N$\ and $S\;$ electrodes connected by a superconducting constriction $
s\;$of length $d$. The dimensions of the constriction transverse to the current
direction  are assumed to be small in comparison with the London
penetration depth in the $S\;$electrode. We consider the system with
diffusive transport, i.e. we suppose that the mean free path $l\;$in the $s\;$
region is small with respect to the constriction length  $d.$
Because of non-conservation of the momentum, the interference of normal
electron wave functions related to reflections from the barriers is
not essential. Nevertheless the coherence of different (ordinary and
Andreev) reflection processes related to the condensate wave function and
nonzero order parameter $\Delta \;$in the superconductor $s\;$is very
important because the inter-barrier distance\ $d\;$is supposed to be small
in comparison with$\;\sqrt{\hbar D/\Delta }$,\ where $D=lv_F/3\;$is the
diffusion coefficient. In what follows we assume that transparencies of both
barriers ${\mathcal D}_{1,2},\;$(averaged over momentum directions) are small
enough to allow the main contribution to the resistance of the system to be due
to the barrier resistances. Tunneling processes determine the dwell times $
\tau _{b1,2}={\mathcal D}_{1,2}v_F/4d\;$in the $s\;$region which are supposed to
be shorter than the inelastic relaxation time $\tau _{in}\;$in the
superconductor $s,\;$so that the following conditions should be fulfilled 
\begin{equation}
\tau _{dif}\ll \tau _{b1,2}\ll \tau _{in}  \label{c1}
\end{equation}
where $\tau _{dif}=\hbar /(D/d^2)\;$is the diffusion time of quasiparticles
through the length $d$\ . It is clear that the quantum nature of the
tunneling processes becomes more pronounced if the tunneling rates $\hbar
/\tau _{b1,2}\;$are comparable with the characteristic scale of the
quasiparticle energy, 
\begin{equation}
\hbar /\tau _{b1,2}\sim \Delta ,  \label{c2}
\end{equation}
because under this condition the classical notion of a quasiparticle whose
dwell time should be longer than its energy, loses its sense. Nevertheless
the Green's function approach enables one to obtain the quantum kinetic
equations as given in \cite{LO} which are valid beyond the classical
limits, i.e. when the quasiparticle energy is not large compared to the
tunneling rate. Note that under the conditions (\ref{c1})
the proximity effect, i.e. the influence of $S\;$and $N\;$electrodes on
the condensate wave function and on the order parameter in the $s$ region, is
strong. We also note that unusual features of transport properties are due to the
significant role of Andreev reflection processeses in the considered system.
These processeses occur in the presence of two potential barriers (at $x=0$
and $x=d$) and the superconducting order parameter which has a two-step
form:\ $\Delta (x)=\Delta \theta (x)\theta (d-x)+\Delta _S$ $\theta
(x-d)\exp (i\varphi )$, where $\varphi \;$is the phase difference between
the superconductors arising at non-zero voltage, $V$ and $\theta (x),$ is the Heavyside
function. As a result, the energy dependent transmission coefficient ${\mathcal D}
_\epsilon (\Delta ,\varphi )\;$ of quasiparticles with energy $\epsilon<\Delta _S$, that
determines the current at low temperatures, appears to be a function
strongly dependent on $V$ through the voltage dependence of $\Delta $
and $\varphi $. All these circumstances result in nontrivial features
of the quasiparticle phase coherent transport through the $N-s-S\;$system
which will be investigated in this paper. It should be noted that some of
these phenomena have been studied in \cite{Z}. We therefore investigate the
transport phenomena in more detail with emphasis upon the case of a weak pairing electron 
interaction in the $s$ region, i.e. when the critical temperature,$\;T_{c0}$ of the
superconductor,$s$ in the absence of the pair-breaking and proximity effect ($\tau
_{b1,2}=\infty $), is small in comparison with the critical temperature of the $S$ electrode,
$\;T_{cS}$. It will be shown that in spite of a small 
ratio $t_c=T_{c0}/T_{cS}\;$the properties of the $N-s-S\;$system\
may radically differ from  properties of a two
barrier $N-N-S\;$structure well studied in the limit$\;t_c=0$ \cite{VZK,r11}. We also study 
the zero bias conductance as a function of
the phase difference between the $S$ electrodes in a
quasiparticle tunneling interferometer with two superconducting electrodes 
coupled by a mesoscopic superconductor $s$. It is
shown that the amplitude of the conductance oscillation may exceed the
conductance of this structure in the normal state.
%page2
\subsection{The N-s-S system}
We consider the $N-s-S\;$system shown in Fig.1a.
As in Refs.\cite{VZK,Z,VZ,Z1,N,r10,Spiv,r11,r12}  we use the approach
based on the equations for the quasiclassical Green's function
$\check{G}=\check{G}({\bf r,p}_F;\epsilon )\;$ which is the 4x4
supermatrix(see Ref.\cite{LO}),  
\[
\check{G}=\left( 
\begin{tabular}{ll}
$\hat{G}^R$ & $\hat{G}^K$ \\ 
$\hat{0}$ & $\hat{G}^A$
\end{tabular}
\right) 
\]
consisting of the retarded $\hat{G}^R\;$,advanced $\hat{G}^A,\;$and Keldysh $
\hat{G}^K,\;$Green's functions which are 2x2 matrices in Nambu space. Note
that we suppose that a stationary solution realizes and the
Green's functions do not depend on time. This non-obvious assumption is justified 
by the final result. The matrix $\hat{G}^K\;$is related to the\ matrix
distribution function  ${\it \hat{f}\;=}$ ${\it f}_0$ $\hat{1}$ $+{\it f}
\hat{\sigma}_z$ 
\begin{equation}
\hat{G}^K=\hat{G}^R{\it \hat{f}-\hat{f}}\hat{G}^A  \label{GK}
\end{equation}
The matrices $\hat{G}^{R,A}\;$have the following form 
\[
\hat{G}^\mu =g^\mu \hat{\sigma}_z+\hat{f}^\mu ,\;\;\hat{f}^\mu =f^\mu i\hat{%
\sigma}_y\exp (i\hat{\sigma}_z\chi ) 
\]
where $\chi \;$is the phase of the order parameter $\mu = R(A)$. The current in the
system is given by the following relation 
\begin{equation}
I=\frac{\sigma {\it A}}8\mbox{Tr}\hat{\sigma}_z\int d\epsilon (\hat{G^R}
\partial _x\hat{G}^K+\hat{G}^K\partial _x\hat{G}^A)  \label{I}
\end{equation}
where ${\it A}=w_yw_z$\ is the cross-section area of the $s\;$region.
The transverse dimensions $w_{y,z}\;$should be
small compared to the London penetration depth. Therefore we need to solve a
one-dimensional equation in the $s\;$region\ ($0<x<d,\;x$-axis coincides
with the direction of the current), where in the considered diffusive case
the matrix $\check{G}\equiv \check{G}%
(x,\epsilon )$ averaged over the momentum direction obeys the equation 
(see Ref.\cite{LO}) 
\begin{equation}
D\partial _x(\check{G}\partial _x\check{G})+i[\epsilon \check{\sigma}_z+%
\check{\Delta},\check{G}]=\check{0}.  \label{E1}
\end{equation}
$\check{\sigma}_z=\check{1}\hat{\sigma}_z\;$is the Pauli supermatrix, and
the order parameter supermatrix in the film is $\check{\Delta}=\check{1}\hat{%
\Delta}\;$where 
\[
\hat{\Delta}=\left( 
\begin{tabular}{ll}
$0$ & $\Delta $ \\ 
$-\Delta ^{*}$ & $0$%
\end{tabular}
\right) 
\]
The order parameter is given by the self-consistency relation which in the
framework of the weak coupling theory has the form 
%page3
\begin{equation}
\hat{\Delta}=\lambda \int_0^{\omega _D}d\epsilon (\hat{f}^R{\it \hat{f}-\hat{%
f}}\hat{f}^A)  \label{self}
\end{equation}
where the constant $\lambda \;$determines the critical temperature $T_{c0}\;$%
of the superconductor $s$ in the absence of pair-breaking factors and the
proximity effect, 
\[
T_{c0}=1.14\omega _D\exp (-1/\lambda )
\]
\ The matrix $\check{G}\;$obeys the normalization condition 
\begin{equation}
\check{G}^2=1  \label{nc}
\end{equation}
$\;$ In order to solve Eq.(\ref{E1})\ we need to take into account the boundary
conditions \cite{r13} that in the diffusive case reduce to \cite{Z2} (see also \cite{LR}).%% 
\begin{equation}
D(\check{G}\partial _x\check{G})(+0)=\epsilon _{b1}d[\check{G}(+0),\check{G}%
_N],\;D(\check{G}\partial _x\check{G})(d-0)=\epsilon _{b2}d[\check{G}_S,%
\check{G}(d-0)]  \label{bc}
\end{equation}
where $\epsilon _{bj}=\rho D/2dR_{bj\Box },\;R_{b1,2\Box }\;$is the
interface resistance per unit area at the $N/s\;(x=0)\;$and\ $s/S\;(x=d)$\
interfaces,\ $\check{G}_{S,N}\;$equilibrium Green's functions in the
electrodes, $\rho \;$is the normal state specific resistivity of the
superconductor\ $s\;$. Note that the energies, $\epsilon _{bj}\;$, are connected
with the characteristic dwell times: $\tau _{bj}=\hbar /\epsilon _{bj}.\;$%
In terms of the Thouless energy $D/d^2=E_{Th}$,the conditions (\ref
{c1})\ may be written  
\begin{equation}
\hbar /\tau _{in}\ll \epsilon _{bj}\ll E_{Th}  \label{c1a}
\end{equation}
Suppose that the length $d$ of the $s\;$region is small enough i.e.$\;d\ll 
\sqrt{\hbar D/\Delta _S}.\;$Then the solution of Eq.(\ref{E1}) is
readily found(see Appendix I).
The retarded and advanced Green's functions are given by 
\begin{equation}
\hat{G}^\mu (\epsilon )=g^\mu (\epsilon )\hat{\sigma}_z+\hat{f}^\mu
(\epsilon )=\frac{\epsilon ^\mu (\epsilon )\hat{\sigma}_z+[\Delta i\hat{\sigma
}_y+i\epsilon _{b2}\hat{f}_S^\mu (\epsilon )]}{\zeta ^\mu (\epsilon )}
\label{Gm}
\end{equation}
where 
\begin{eqnarray}
\zeta ^\mu (\epsilon ) &=&[(\epsilon ^\mu (\epsilon ))^2-\Delta ^2-2\Delta
\i\epsilon _{b2}f_S^\mu \cos \varphi +(\epsilon _{b2}f_S^\mu )^2]^{1/2}\nonumber \\
\epsilon ^{R,A}(\epsilon ) &=&\epsilon +i\epsilon _{b2}g_S^{R,A}(\epsilon )
\pm i\epsilon _{b1}\nonumber
\end{eqnarray}
The Keldysh function is given by Eq.(\ref{GK2}) and it is convenient to 
separate the anomalous part $\hat{G}_a^K\;$ to give 
\[
\hat{G}^K=\hat{G}^Rn-n\hat{G}^A+\hat{G}_a^K 
\]
We have for $\hat{G}_a^K\;$
\begin{equation}
\hat{G}_a^K=(\hat{E}_a^K-\hat{G}^R\hat{E}_a^K\hat{G}^A)\frac 1{(\zeta
^R+\zeta ^A)}  \label{GKa}
\end{equation}
with the anomalous self-energy 
\begin{equation}
\hat{E}_a^K=2i\epsilon _{b1}\{n_{-}(\epsilon )+\hat{\sigma}_z[n_{+}(\epsilon
)-n(\epsilon )]\}  \label{EKa}
\end{equation}
where $n(\epsilon )=\tanh (\epsilon /2T),$ 
\begin{equation}
n_{\pm }(\epsilon )=[n(\epsilon +eV)\pm n(\epsilon -eV)]/2
\end{equation}
$\;$Thus the anomalous part $\hat{G}_a^K\;$is determined by $\hat{E}_a^K$
which contains only the self-energy\ depending on the N-electrode\ Green's
function $\hat{G}^K$. Using Eq.(\ref{GKa}) enables one to find the
non-equilibrium part of the matrix distribution function 
\begin{equation}
\delta {\it \hat{f}}=({\it f}_0-n)+{\it f}\hat{\sigma}_z  \label{df}
\end{equation}
which determines the anomalous part of $\hat{G}^K$ : 
\begin{equation}
\hat{G}_a^K=\hat{G}^R\delta {\it \hat{f}}-\delta {\it \hat{f}}\hat{G}^A
\end{equation}
%page4
From Eq.(\ref{GKa}) we find for the non equilibrium parts of the distribution
functions 
\begin{equation}
{\it f}=\frac 1{4\nu (\zeta ^R+\zeta ^A)}\mbox{Tr}\hat{E}_a^K(1-\hat{G}^A%
\hat{G}^R)
\end{equation}
\begin{equation}
\delta {\it f}_0=\frac 1{4\nu (\zeta ^R+\zeta ^A)}\mbox{Tr}\hat{E}_a^K(\hat{%
\sigma}_z-\hat{G}^A\hat{\sigma}_z\hat{G}^R)
\end{equation}
where $\nu =
%TCIMACRO{\func{Re}}
%BeginExpansion
\mathop{\rm Re}
%EndExpansion
g^R\;$is the density of states, $\delta {\it f}_0={\it f}_0-n,$.
Using Eq.(\ref{EKa}) for $\hat{E}_a^K$ the
non-equilibrium part of the distribution functions may be written in the following form 
\begin{eqnarray}
\delta {\it f}_0 &=&a_{+}(n_{+}-n)+bn_{-}  \label{dfa} \\
{\it f} &=&a_{-}n_{-}-b(n_{+}-n)  \nonumber
\end{eqnarray}
where 
\[
a_{\pm }=\frac{\epsilon _{b1}M_{\pm }}{2\nu 
%TCIMACRO{\func{Im} }
%BeginExpansion
\mathop{\rm Im}
%EndExpansion
\zeta ^R},\;\;b=\frac{\Delta \epsilon _{b1}\epsilon _{b2}}{\nu 
%TCIMACRO{\func{Im} }
%BeginExpansion
\mathop{\rm Im}
%EndExpansion
\zeta ^R}\frac{%
%TCIMACRO{\func{Re} }
%BeginExpansion
\mathop{\rm Re}
%EndExpansion
f_S^R}{\left| \zeta ^R\right| ^2}\sin \varphi  
\]
\[
M_{\pm }=1-g^Rg^A\pm [\Delta ^2-2\epsilon _{b2}\Delta 
%TCIMACRO{\func{Im} }
%BeginExpansion
\mathop{\rm Im}
%EndExpansion
f_S^R\cos \varphi +(\epsilon _{b2}\left| f_S^R\right| )^2]\frac 1{\left|
\zeta ^R\right| ^2} 
\]
We took into account that $\zeta ^A=-(\zeta ^R)^{*},\;$ $g^A=-(g^R)^{*}.$
From the self-consistency relation (\ref{self}) we obtain the following
system of equations for $\Delta \;$and\ $\varphi $, 
\begin{equation}
\Lambda \Delta =\epsilon _{b2}(\alpha \cos \varphi -\beta _1\sin \varphi ),
\label{sc1}
\end{equation}
\begin{equation}
\beta _0\Delta =\epsilon _{b2}(\alpha \sin \varphi +\beta _1\cos \varphi ),
\label{sc2}
\end{equation}
where 
\[
\Lambda =\ln (T/T_{c0})-\int_0^\infty d\epsilon \left( {\it f}_0(\epsilon)
%TCIMACRO{\func{Re} }
%BeginExpansion
\mathop{\rm Re}
%EndExpansion
\frac 1{\zeta ^R(\epsilon )}-\frac{n(\epsilon )}\epsilon \right) 
\]
\[
\alpha =-\int_0^\infty d\epsilon {\it f}_0(\epsilon )%
%TCIMACRO{\func{Im} }
%BeginExpansion
\mathop{\rm Im}
%EndExpansion
\frac{f_S^R(\epsilon )}{\zeta ^R(\epsilon )}
\]
\[
\beta _k=\int_0^\infty d\epsilon {\it f}(\epsilon )%
%TCIMACRO{\func{Im} }
%BeginExpansion
\mathop{\rm Im}
%EndExpansion
\frac{k-1+kif_S^R(\epsilon )}{\zeta ^R(\epsilon )}\;,\;k=0,1.
\]
Note that Eq.(\ref{sc2}) is the consequence of the current
conservation law in the $x$-direction.\ Introducing the normalized order
parameter $\delta =\Delta /\epsilon _{b2}\;$one can reduce Eqs.
(\ref{sc1}) and (\ref{sc2}) to the equivalent ones 
\begin{equation}
\delta =\sqrt{\frac{\alpha ^2+\beta _1^2}{\Lambda ^2+\beta _0^2}},
\label{sc1a}
\end{equation}
\begin{equation}
\exp (i\varphi )=\frac{\alpha \Lambda +\beta _0\beta _1+i(\alpha \beta
_0-\Lambda \beta _1)}{\sqrt{\alpha ^2+\beta _1^2}\sqrt{\Lambda ^2+\beta _0^2}%
}  \label{sc2a}
\end{equation}
From the self-consistency equations (\ref{sc1}) and (\ref{sc2}) or 
(\ref{sc1a}) and (\ref{sc2a}), it follows that the stationary solution for the order
parameter exists at arbitrary $V$\ and transition to the ac Josephson effect (time
dependent phase difference\ $\varphi $) does not occur with increasing voltage.
In other words the critical current of the $S/s\;$tunnel junction is absent in
the considered mesoscopic system. Such a situation differs radically from that
in a single $S/S\;$tunnel junction composed of two bulk
superconductors. If at least one of the two superconductors has mesoscopic 
dimensions, it is important how it is connected with the conductors and
%page5
non-equilibrium states arising in the presence of the current 
play a significant role in this case.
In our system one of the important aspects of the
non equilibrium state is the quasiparticle charge-imbalance determined by
the distribution function ${\it f}(\epsilon )\;$, and, as a consequence, the
gauge-invariant potential $\mu =\Phi + (\hbar/2e) \partial _t\chi $\ 
in the $s$ region, where $\Phi \;$is electrical potential and $\chi $ is the
order parameter phase. Under the assumption (\ref{c1}) the solution for the
phase difference between the superconductors is stationary for arbitrary
voltages. Therefore we can set $\chi =0$ in the $s$ region so that $\mu =\frac 1e%
\int_0^\infty d\epsilon {\it f}(\epsilon )\nu (\epsilon )\;$coincides with
the voltage between the superconductors $\Phi \neq -(\hbar/2e) \partial
_t\varphi =0.$ In other words the Josephson relation between the frequency 
(equal to zero)\ and the voltage drop across the superconducting tunnel junction
is violated in the structure under consideration.
The current may be calculated at any point\ $x$ and, for example, at
the $N/s$ interface we obtain  
\begin{equation}
I=\frac 1{2eR_N}\int_{-\infty }^\infty d\epsilon \{F_{-}(\epsilon
)n_{-}(\epsilon )+F_{+}(\epsilon )[n_{+}(\epsilon )-n(\epsilon )]\}\;,
\label{Cur}
\end{equation}
where 
\begin{eqnarray*}
F_{-}(\epsilon ) &=&(1+r)\nu (\epsilon )[1-a_{-}(\epsilon
)],\;F_{+}(\epsilon )=\nu (\epsilon )b(\epsilon )(1+r)\;, \\
r &=&R_{b2}/R_{b1}=\epsilon _{b1}/\epsilon _{b2}\;.
\end{eqnarray*}
$\;$If the $S\;$electrode is a conventional BCS superconductor, 
\[
g_S^R(\epsilon )=f_S^R(\epsilon )\epsilon /\Delta _S=\epsilon /\sqrt{%
(\epsilon +i0)^2-\Delta _S^2}\;. 
\]
Then for $\left| \epsilon \right| <\Delta _{S,\;}b(\epsilon
)=0,\;F_{+}(\epsilon )=0\;$and at $\left| eV\right| <\Delta \;$at zero
temperature the current reads 
\begin{equation}
I=\frac 1{eR_N}\int_0^Vd\epsilon F_{-}(\epsilon )\;.  \label{Cur0}
\end{equation}
Note that the function $F_{-}(\epsilon )=F_{-}(\epsilon ;V)\;$depends on
voltage through the voltage dependence of $\Delta $\ and $\varphi .\;$It
represents the transmission coefficient of the system which determines the
efficiency of Andreev reflection processes. Taking into account that $%
b(\epsilon )=0$ and assuming $\left| eV\right| <\Delta _S\;$, we find for
the non-equilibrium part of the distribution functions 
\begin{eqnarray}
{\it f}(\epsilon ) &=&a_{-}(\epsilon )\mbox{sgn}(eV)\theta (\left| eV\right|
-\left| \epsilon \right| )\;, \\
\delta {\it f}_0(\epsilon ) &=&-a_{+}(\epsilon )\mbox{sgn}(\epsilon )\theta
(\left| eV\right| -\left| \epsilon \right| )\;.\;
\end{eqnarray}
Consider the case of small critical temperatures of the superconductor $%
s,\;T_{c0}/T_{cS}\ll 1,\;$and also assume that the following condition is
fulfilled 
\begin{equation}
\Delta ,\;\epsilon _{b1},\;\epsilon _{b2},eV\ll \Delta _S\;.  \label{c3}
\end{equation}
In this case Eq.(\ref{sc1a}) and (\ref{sc2a})\ for $\Delta $ and $\varphi $
can be simplified and presented in the form (see Appendix II). 
\begin{equation}
\delta =\sqrt{\frac{\alpha ^2+\beta _0^2}{\Lambda ^2+\beta _0^2}},
\label{sc3a}
\end{equation}
\begin{equation}
\cos \varphi +i\sin \varphi =\frac{\alpha \Lambda -\beta _0^2+i\beta
_0(\alpha +\Lambda )}{\sqrt{\alpha ^2+\beta _0^2}\sqrt{\Lambda ^2+\beta _0^2}%
}\;.  \label{sc3b}
\end{equation}
From Eq.(\ref{Cur0}) and (\ref{a-}) we find the current 
\begin{equation}
\frac I{(\epsilon _{b1}+\epsilon _{b2})/eR_N}=2\Omega _\varphi \int_0^vdu%
\frac{\nu (u,\Omega _\varphi )}{u^2+r^2+\Omega _\varphi +\left| \zeta
(u,\Omega _\varphi )\right| ^2}\;.  \label{ncur}
\end{equation}
The I-V curves obtained by numerical calculations results on the basis of
Eqs.(\ref{sc3a}),\ (\ref{sc3b}) and (\ref{ncur}) are presented in Fig.2. One
can see that (as a consequence of the order parameter suppression in the $%
s\; $region) the differential conductance becomes negative with growing
voltage.

In general the solution of Eqs.(\ref{sc3a}),\ (\ref{sc3b}) and the current
can only be determined numerically because the formulas are rather
complicated. Nevertheless the zero-bias conductance ${\it g}_0={\it G}(0)/%
{\it G}_N\;$ can be found from Eq.(\ref{ncur}), where\ ${\it G}(V)=dI/dV\;$%
.\ It is given by 
\begin{equation}
{\it g}_0(\delta ,r)=\frac{(1+r)r(1+\delta )^2}{[r^2+(1+\delta )^2]^{3/2}}
\label{g0}
\end{equation}
where according to Eqs.(\ref{sc3a}) and (\ref{sc1})$\;\delta =\Delta
/\epsilon _{b2}\;$is defined\ by the equation 
\begin{equation}
(\delta +1)\ln \frac{[r+\sqrt{r^2+(\delta +1)^2}]}{\delta _0}=\ln \frac 4{t_c%
}  \label{D0}
\end{equation}
It follows from Eq.(\ref{D0})\ that (under the condition (\ref{c3})) the
proximity effect is strong, i.e. $\Delta \gg \Delta _{0.}\;$Moreover 
\[
\frac \Delta {\Delta _0}\rightarrow \infty \text{\ at}\;\Delta _0\rightarrow
0\;, 
\]
i.e. due to the proximity effect, anomalously big enhancements of the order
parameter occur at very weak pairing electron interaction in the $s\;$%
region.\ Assuming $\delta \ll 1$, one can obtain from (\ref{D0})\ that 
\begin{equation}
\frac \Delta {\Delta _0}=\frac 1{\delta _0}\frac{\ln \frac{4\Delta _S}{%
\epsilon _{b2}(r+\sqrt{r^2+1})}}{\ln \frac{(r+\sqrt{r^2+1})}{\delta _0}}\;.
\label{asr}
\end{equation}
This expression is valid for very small $\delta _0\;$which satisfies the
condition 
\[
\ln \frac{(r+\sqrt{r^2+1})}{\delta _0}\gg \ln \frac{4\Delta _S}{\epsilon
_{b2}(r+\sqrt{r^2+1})}\; 
\]
that is fulfilled provided $\delta _0\ll (\epsilon _{b1}+\epsilon
_{b2})^2/\Delta _S^2\ll 1.\;$In particular if $\epsilon _{b1}+\epsilon
_{b2}\sim 10^{-2}\Delta _S$ the requirement $\delta _0\ll 10^{-4\;}$means
that (\ref{asr})\ is valid provided $T_{c0}\;$is anomalously small, $%
T_{c0}\ll 10^{-6}T_{cS},\;$then $\Delta \gg 10^4\Delta _0.\;$ It can be seen
from condition (\ref{c3}) that one can ignore the presence of the order
parameter in the $s$\ region ($\delta \ll 1)\;$only if the pairing
interaction in the s region is very weak, i.e.\ $T_{c0}\ll \epsilon
_{b2}(\epsilon _{b1}+\epsilon _{b2})^2/\Delta _S^2$ . The zero-bias
conductance as a function of {\it $t_c$}\ is shown in Fig.3 for different
parameters {\it $r=R_{b2}/R_{b1}\;$(}and{\it \ }$\epsilon _{b2}=0.05\Delta
_S $)$.$ One can see that the normalized conductance may be both smaller and
bigger than unity. In particular\ from (\ref{g0}) we find that for $r>1/%
\sqrt{2}\;$the maximum value of the conductance, ${\it g}$ corresponds to $%
\delta =\delta _m$\ where $(1+\delta _m)=\sqrt{2}r\;$and Eq.(\ref{g0})
equals 
\[
{\it g}_{0\max }=\frac 2{3\sqrt{3}}(1+r)\;. 
\]
From Eq.(\ref{D0}) we find that the maximum conductance is realized for the
case when the critical temperature $T_{c0}=T_{c0}^m,$ where 
\begin{equation}
T_{c0}^m=4T_{cS}\left[ \frac{\epsilon _{b1}(1+\sqrt{3})}{4T_{cS}}\right]
^{1/(1-1/\sqrt{2}r)}\;.  \label{Tcm}
\end{equation}
Eq.(\ref{Tcm}) is applicable for $r\;$satisfying the condition $T_{c0}^m\ll
T_{cS};$ in particular it is true for $r>3\sqrt{3}/2-1\;$which corresponds
to ${\it g}_{0\max }>1.$
At low temperatures\ $T\ll \Delta _S\;$for zero-bias conductance we find
from (\ref{Cur}) 
\begin{equation}
{\it g}(t)=2(1+r)\Omega (t)\int_0^\infty \frac{du}{\cosh ^2u}\frac{\mbox{Re}%
(2tu+ir)/\zeta (2tu,\Omega (t))}{(2tu)^2+r^2+\Omega (t)+\left| \zeta
(2tu,\Omega (t))\right| ^2}  \label{g(t)}
\end{equation}
where $t=T/\epsilon _{b2},\;\Omega (t)=(1+\delta (t))^2\;$and\ $\zeta
(u,\Omega )=[(u+ir)^2-\Omega ]^{1/2}$, the function $\delta (t)\;$is defined
by the equation 
\begin{equation}
\delta =\frac{\alpha (\Omega ,t)}{\Lambda (\Omega ,t)}\;  \label{d(t)}
\end{equation}
with 
\[
\alpha (\Omega ,t)=\ln \frac{4\Delta _S}{\left( \sqrt{\Omega +r^2}+r\right)
\epsilon _{b2}}-\int_0^\infty du\frac 2{\exp u+1}\mbox{Re}\frac 1{\zeta
(tu,\Omega )} 
\]
\[
\Lambda (\Omega ,t)=\ln \frac{\sqrt{\Omega +r^2}+r}{\delta _0}+\int_0^\infty 
\frac{du}{\cosh ^2u}\ln \frac{\left| \zeta (2tu,\Omega )+2ut+ir\right| }{%
\sqrt{\Omega +r^2}+r} 
\]
The results of numerical calculations on the basis of Eqs.(\ref{g(t)}),\ (%
\ref{d(t)})\ are presented in Figs. 4 and 5.\ We see that the conductance may be a
non-monotonic function of temperature that radically differs from the
corresponding dependencies occurring in the case of normal mesoscopic region
with $T_{c0}=0\;$shown by\ dashed lines in Figs. 4 and 5.\ Thus a weak
pairing electron interaction results in significant qualitative (for $r\geq
1 $)\ and quantitative changes of the conductance dependence with respect to
the case of structure within the normal mesoscopic region. Fig.6 shows that if 
$r<1,\;$the conductance may be non-monotonic function of temperature even at 
$T_{c0}=0.\;$We see that the pairing electron interaction results in a shift
of the position of the conductance maximum to higher temperatures together
with an increase in width of the maximum. The latter is due to the slow
decrease of the order parameter with increasing temperature.

Consider the case when the resistance of the barrier at the $N/s$ interface
is small enough\ ($r\gg 1$). To be more exact we suppose that 
\begin{equation}
\epsilon _{b1}\gg \Delta \;,\;\epsilon _{b2}\;.
\end{equation}
i.e.$\;r\gg \delta .$ In this case the energy gap is absent in the
superconductor $s$ due to a strong pair-breaking effect of the normal
electrode. If the condensate Green's function is small, all the expressions
are significantly simplified and from (\ref{ncur}) at $T,eV\ll \Delta _S\;$%
we find for the current 
\begin{equation}
I=\frac{(\epsilon _{b1}+\epsilon _{b2})}{eR_N}(\delta ^2+2\delta \cos
\varphi +1)%
%TCIMACRO{\func{Im} }
%BeginExpansion
\mathop{\rm Im}
%EndExpansion
\Psi (\Gamma +ieV/2\pi T)\;,  \label{I(V,T)}
\end{equation}
where $\Psi (z)\;$is the digamma-function, $\Gamma =1/2+\epsilon _{b1}/2\pi
T,\;\delta \;$and cos$\varphi \;$are given by Eqs.(\ref{sc3a}),\ (\ref{sc3b}%
)\ with 
\begin{equation}
\Lambda =\ln (T/T_{c0})+%
%TCIMACRO{\func{Re} }
%BeginExpansion
\mathop{\rm Re}
%EndExpansion
\Psi (\Gamma +ieV/2\pi T)-\Psi (1/2),\;\;\beta _0=%
%TCIMACRO{\func{Im} }
%BeginExpansion
\mathop{\rm Im}
%EndExpansion
\Psi (\Gamma +ieV/2\pi T),  \label{coef}
\end{equation}
\[
\alpha =\Delta _S\int_0^{\Delta _S}d\epsilon \frac{\epsilon n_{+}(\epsilon )%
}{(\epsilon ^2+\epsilon _{b1}^2)\sqrt{\Delta _S^2-\epsilon ^2}}\;. 
\]
At zero temperature we obtain from Eqs.(\ref{I(V,T)}),\ (\ref{coef}) 
\begin{equation}
\frac I{(\epsilon _{b1}+\epsilon _{b2})/eR_N}=(\delta ^2+2\delta \cos
\varphi +1)\arctan (v/r)\;,
\end{equation}
where Eqs.(\ref{coef})\ reduce to 
\[
\Lambda =\ln \frac{2\sqrt{v^2+r^2}}{\delta _0},\;\;\alpha =\ln \frac{2\delta
_0}{t_c\sqrt{v^2+r^2}}\;,\;\beta _0=\arctan (v/r)\;. 
\]
The I-V curves computed for this case are shown in Fig.7. At small voltages
we have $IR_N={\it g}_0V\;$, where the normalized conductance is given by
the expression 
\[
{\it g}_0=\frac{\ln ^2\frac 4{t_c}}{r\ln ^2\frac{2r}{\delta _0}}\;. 
\]
At large voltages, $\epsilon _{b1}\ll eV\ll \Delta _S,\;$the normalized
current has the form 
\[
\frac I{(\epsilon _{b1}+\epsilon _{b2})/eR_N}=\frac{\ln ^2(2\Delta
_S/eV)+\ln ^2(2eV/\Delta _0)+2\sqrt{[\ln ^2(2\Delta _S/eV)+1][\ln
^2(2eV/\Delta _0)+1]}}{\ln ^2(2eV/\Delta _0)+1}\;. 
\]
At $(2eV)^2>\Delta _S\Delta _0$ this function slowly decreases with
increasing voltage.
\subsection{Quasiparticle interferometer\ }
Consider a quasiparticle interferometer composed of three tunnel junctions
(see Fig.1b) in which the phase difference $\varphi \;$between two different 
$S/s\;$interfaces is set by an external magnetic field. A similar system in
which $S$\ and $N$ electrodes were in contact with a normal metal was
considered in \cite{VZ,Z1,N,r10}. Suppose that two barriers at the $S/s$
interfaces are symmetrical with resistances equal to $R_{b2}$\ and the
resistance of the barrier at the $N/s$ interface equals $R_{b1}.\;$We again
assume that the resistance of the system is determined by the barriers and
in the normal state is given by the expression $R_N=R_{b2}/2+R_{b1}$.
Assuming that the width of the superconductor $s$ is small, $W\;\ll \sqrt{%
\hbar D/\Delta }$\ \ one can neglect the spatial\ variation of the Green's
function. Then Eq.(\ref{A1}) is valid with 
\begin{equation}
\check \Sigma =i\epsilon _{b2}\check G_{S+}+i\epsilon _{b2}\check G%
_{S-}+\epsilon _{b1}\check G_N\;,  \label{Sig}
\end{equation}
where the Green's functions $\check G_{S\pm }\;$correspond to the phases $%
\pm \varphi /2,\;\epsilon _{bj}=\rho Dw_j/2dWR_{bj\Box }\;,w_1=W$\ being the
width of the $s$ region and $w_2$ is the width of the $S/s$ interfaces. As
in previous cases we find 
\begin{equation}
\hat G^\mu (\epsilon )=g^\mu (\epsilon )\hat \sigma _z+\hat f^\mu (\epsilon
)=\frac{\epsilon ^\mu (\epsilon )\hat \sigma _z+\Delta ^\mu (\epsilon )i\hat 
\sigma _y}{\zeta ^\mu (\epsilon )}\;,  \label{Gm1}
\end{equation}
where 
\begin{eqnarray}
\zeta ^\mu (\epsilon ) &=&\{(\epsilon ^\mu (\epsilon ))^2-[\Delta ^\mu
(\epsilon )]^2\}^{1/2}\;,  \nonumber \\
\epsilon ^{R,A}(\epsilon ) &=&\epsilon +2i\epsilon _{b2}g_S^{R,A}(\epsilon
)\pm i\epsilon _{b1},\;\Delta ^\mu (\epsilon )=\Delta +2i\epsilon
_{b2}f_S^\mu (\epsilon )\cos (\varphi /2)\;.  \nonumber
\end{eqnarray}
From the self-consistency relation (\ref{self})\ at zero voltage between the 
$S\;$and $N$ electrodes the following system of equations for $\Delta \;$%
and\ $\varphi $ can be found 
\begin{equation}
\Lambda \Delta =\alpha =-2\epsilon _{b2}\cos (\varphi /2)\int_0^\infty
d\epsilon n(\epsilon )%
%TCIMACRO{\func{Im} }
%BeginExpansion
\mathop{\rm Im}
%EndExpansion
\frac{f_S^R(\epsilon )}{\zeta ^R(\epsilon )}\;.  \label{sc1aa}
\end{equation}
where 
\[
\Lambda =\ln (T/T_{c0})-\int_0^\infty d\epsilon \left( 
%TCIMACRO{\func{Re}}
%BeginExpansion
\mathop{\rm Re}
%EndExpansion
\frac 1{\zeta ^R(\epsilon )}-\frac 1\epsilon \right) n(\epsilon ) 
\]
At $T=0$ assuming as before that $\Delta _0,\epsilon _{b1,2}\ll \Delta _S\;$%
, we obtain 
\begin{equation}
\Lambda =\ln \frac{\sqrt{\bar \Delta _\varphi ^2+\epsilon _{b1}^2}+\epsilon
_{b1}}{\Delta _0}\;,\;  \label{Laa}
\end{equation}
\begin{equation}
\alpha =2\epsilon _{b2}\cos (\varphi /2)\ln \frac{4\Delta _S}{\epsilon _{b1}+%
\sqrt{\epsilon _{b1}^2+\bar \Delta _\varphi ^2}}\;\;,  \label{ao}
\end{equation}
where $\bar \Delta _\varphi =\Delta +2\epsilon _{b2}\cos (\varphi /2).$\ It
is convenient to introduce the function $\bar \delta _\varphi :$ $\Delta
/2\epsilon _{b2}=\bar \delta _\varphi \cos (\varphi /2)$, then from Eqs.(\ref
{sc1aa}) and (\ref{ao})\ the following equation for $\bar \delta _\varphi $
can be found 
\begin{equation}
(\bar \delta _\varphi +1)\ln \frac{\sqrt{(\bar \delta _\varphi +1)^2\cos
^2(\varphi /2)+r^2}+r}{\delta _0}=\ln \frac 4{t_c}\;,
\end{equation}
where $\delta _0=\Delta _0/2\epsilon _{b2}.\;$After calculations similar to
those carried out in Refs.\cite{VZ,Z1,N,r10} we obtain for the zero-bias
conductance of a symmetrical quasiparticle interferometer at zero
temperature 
\begin{equation}
\frac{{\it G}(0,\varphi )}{{\it G}_N}=\frac{(1+r)r(1+\bar \delta _\varphi
)^2\cos ^2(\varphi /2)}{[r^2+\cos ^2(\varphi /2)(1+\bar \delta _\varphi
)^2]^{3/2}}\;.  \label{G0f}
\end{equation}
Note that at $\varphi =0$ Eq.(\ref{G0f})\ is identical to Eq.(\ref{g0}).\
Thus the amplitude of the conductance oscillations may exceed ${\it G}_N\;$%
if the ratio $r$\ is\ large enough.\ For $r^2\gg (\delta +1)^2\;$(when the
energy gap is absent in the $s$ region), Eq.(\ref{G0f}) yields 
\begin{equation}
\frac{{\it G}(0,\varphi )}{{\it G}_N}=\frac{\ln ^2(4/t_c)}{r\ln ^2(2r/\delta
_0)}\cos ^2(\varphi /2)\;.  \label{Gf}
\end{equation}
Since $4/t_c\gg 2r/\delta _0\gg 1\;(\Delta _S\gg \epsilon _{b1})$ the\
amplitude of the conductance oscillations appears to be much larger than in
the case of interferometer with a normal mesoscopic region \cite{VZ,Z1,N,r10}

\subsection{Conclusion}

In conclusion, we have studied phase-coherent diffusive transport through
different tunnel structures with $S$\ and $N$\ electrodes coupled by a
mesoscopic superconductor $s$. Our study has centered upon the case of a
weak pairing electron interaction in the $s$\ region which defines a
critical temperature $T_{c0}$\ $\ll T_{cS}.$\ If the dwell time in the $s$
region, determined by the tunneling processes $\tau _b$, is small or
comparable with\ $\hbar /\Delta _0$, the proximity effect is strong, i.e.
the order parameter $\Delta \gg \Delta _0\sim T_{c0}\;$. As a consequence,
the subgap conductance of an $N-s-S\;$tunneling structure and of a
quasiparticle tunneling interferometer, depends strongly upon the pairing
electron interaction in the $s\;$region when $T_{c0}\ll T_{cS}\;$. Depending
upon the ratio of the barrier resistances, the value of the subgap
conductance, determined by Andreev reflection processes may be both larger
and smaller than the conductance of these structures in the normal state. We
have shown that even weak pairing electron interaction may result in the
significant qualitative (in the case $r\geq 1$)\ and quantitative change of
the conductance temperature dependence with respect to the case of
structures with the normal mesoscopic region. The subgap current
non-monotonously depends upon the voltage, due to the suppression of the
order parameter in the mesoscopic superconductor.\\ 
\par
{\bf Acknowledgments. }The authors (A.V.Z., A.F.V.) acknowledge the
financial support of CRDF project RPA-165,\ A.V.Z. acknowledge the financial
support of Royal Society and Russian Fond For Fundamental Research (project
96-02-18613) A.V.Z. is grateful to C.J. Lambert for hospitality during his
visit to Lancaster where this work has been completed.
\newpage
\section{appendix}
\renewcommand{\theequation}{A\arabic{equation}}
\setcounter{equation}{0}
Integrating Eq.(\ref{E1})\ over$\;x$ and taking into account the boundary 
conditions, we obtain the following equation for $\check{G}\;$(in the
following $\check{G}\;$denotes the function averaged over the length $d$ ) 
\begin{equation}
\lbrack \check{E},\check{G}]=\check{0},  \label{A1}
\end{equation}
where 
\[
\check{E}=\epsilon \check{\sigma}_z+\check{\Delta}+\check{\Sigma}, 
\]
\begin{equation}
\check{\Sigma}=i\epsilon _{b1}\check{G}_N+i\epsilon _{b2}\check{G}_S
\end{equation}
When writing equations (\ref{E1}) and (\ref{A1})\ we disregarded inelastic
collisions due to condition (\ref{c1}).\ Let the
potential of the superconducting electrode be zero and the potential of the
normal electrode be equal to $V$,\ so that 
\begin{equation}
\hat{G}_N^{R,A}=\pm \hat{\sigma}_z,\;\hat{G}_N^K=(1+\hat{\sigma}%
_z)n(\epsilon +eV)+(1-\hat{\sigma}_z)n(\epsilon -eV),
\end{equation}
\begin{equation}
\hat{G}_S^{R,A}=g_S^{R,A}\hat{\sigma}_z+\hat{f}_S^{R,A},\;\hat{G}_S^K=(\hat{G%
}_S^R-\hat{G}_S^A)n(\epsilon ).
\end{equation}
When finding the solution of (\ref{A1}) it is convenient to let the phase of
the order parameter in the $s$ layer equal zero and the phase of the
superconducting electrode $S$ be equal to the phase difference $\varphi \;$%
which arises in the presence of the current. Then from (\ref{A1}) and (\ref
{nc})\ we find the expressions for the retarded and advanced Green's given in 
Eq.(\ref{Gm}).
The equation for $\hat{G}^K\;$has the form 
\begin{equation}
\hat{E}^R\hat{G}^K-\hat{G}^K\hat{E}^A=\hat{G}^R\hat{E}^K-\hat{E}^K\hat{G}^A
\label{EGK}
\end{equation}
where $\hat{E}^K=\hat{\Sigma}_1^K+\hat{\Sigma}_2^K.\;$It is useful to take
into account that 
\begin{equation}
\hat{E}^{R,A}=\zeta ^{R,A}\hat{G}^{R,A}  \label{ERA}
\end{equation}
Then using Eq.(\ref{nc}), we have 
\begin{equation}
\hat{G}^R\hat{G}^K+\hat{G}^K\hat{G}^A=\hat{0}  \label{nc2}
\end{equation}
Therefore from (\ref{ERA}) it follows that $\hat{E}^R\hat{G}^K-\hat{G}^K\hat{%
E}^A=(\zeta ^R+\zeta ^A)\hat{G}^R\hat{G}^K$\ and from (\ref{EGK})\ we find for
the Keldysh functions 
\begin{equation}
\hat{G}^K=(\hat{E}^K-\hat{G}^R\hat{E}^K\hat{G}^A)\frac 1{(\zeta ^R+\zeta ^A)}
\label{GK2}
\end{equation}
\section{appendix}
Here we present simplified formulas for $\Omega\;$,$\alpha\;$ and $\beta\;$ by
taking into account the following identity which may be readily proved for
small energies $\epsilon \ll \Delta _S:$ 
\[
(2\nu 
%TCIMACRO{\func{Im}}
%BeginExpansion
\mathop{\rm Im}
%EndExpansion
\zeta ^R)(u,\Omega _\varphi )=r\frac{(u^2+r^2+\Omega _\varphi )}{\left|
\zeta (u,\Omega _\varphi )\right| ^2}+r, 
\]
Using the notations $u=\epsilon /\epsilon _{b2},\zeta (u,\Omega
)=[(u+ir)^2-\Omega ]^{1/2},\;\Omega _\varphi =\delta ^2+2\delta \cos \varphi
+1\;$ one can obtain the expressions for $a\pm $ from (\ref{dfa}). 
\begin{eqnarray}
a_{+} &=&1,  \label{a+} \\
a_{-} &=&1-\frac{2\Omega _\varphi }{u^2+r^2+\Omega _\varphi +\left| \zeta
(u,\Omega _\varphi )\right| ^2}.  \label{a-}
\end{eqnarray}
At zero temperature and $eV\ll \Delta _S$, one has 
\begin{equation}
\Lambda =\Lambda _0-\int_0^{eV}d\epsilon \delta {\it f}_0(\epsilon )%
%TCIMACRO{\func{Re} }
%BeginExpansion
\mathop{\rm Re}
%EndExpansion
\frac 1{\zeta ^R(\epsilon )},  \label{La1}
\end{equation}
where 
\[
\Lambda _0=\ln \frac{\sqrt{\Omega _\varphi +r^2}+r}{\delta _0},
\]
\[
\delta _0=\Delta _0/\epsilon _{b2},\;\Delta _0=1.76T_{c0},
\]
$\Delta _0\;$being the gap of the superconductor $s$\ at $T=0$ in the
absence of pair-breaking\ factors\ and the proximity effect ($\epsilon _{bj}=0)$.
Introducing the normalized voltage $v=Ve/\epsilon _{b2}$,
from Eq.(\ref{La1}) we find expressions for $\alpha\;$ and $\beta\;$, (see
Eqs.\ref{sc1} and (\ref{sc2}).
\begin{equation}
\Lambda =\ln \frac{\left| \zeta (v,\Omega _\varphi )+v+ir\right| }{\delta _0}
\end{equation}
\[
\alpha =\ln \frac{4\delta _0}{\left| \zeta (v,\Omega )+v+ir\right| t_c}
\]
\[
\beta _0=-\beta _1=\frac r2\int_0^vdu\frac{M_{-}(u,\Omega _\varphi )}{\nu
(u,\Omega _\varphi )\left| \zeta (u,\Omega _\varphi )\right| ^2}
\]

\newpage 
\begin{figure}
\centerline{\psfig{figure=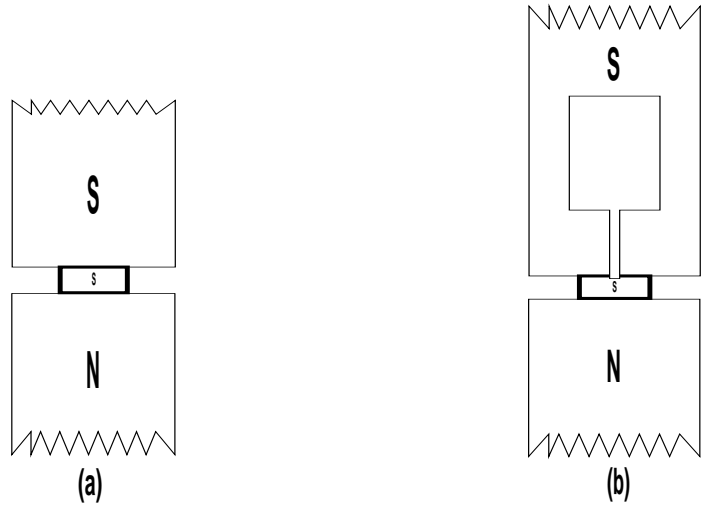,width=10cm,height=8cm}}
\caption{The $N/s/S$ system under consideration. (a). The $S/s$ and the $N/s$\\
interfaces have $R_{b2}$ and $R_{b1}$ barrier resistances respectivly.
\newline (b). The schematical representation of the Andreev interferometer.}
\label{Fig.1}
\end{figure}

\begin{figure}
\centerline{\psfig{figure=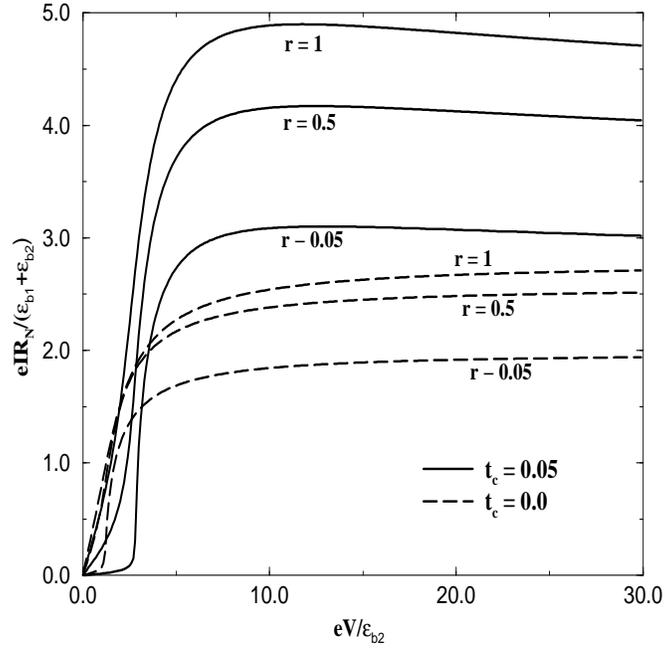,width=10cm,height=10cm}}
\caption{I-V curves at zero temperature for r less than 1,
$\epsilon _{b2}/\Delta_S=0.05.$}
\label{Fig.2}
\end{figure}

\begin{figure}
\centerline{\psfig{figure=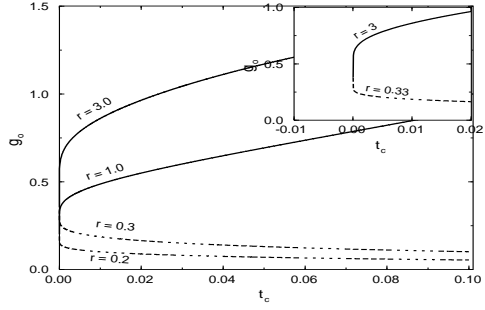,width=8cm,height=8cm}}
\caption{Dependence of the normalised conductance, $g_o$ on 
$t_c$.}
\label{Fig.3}
\end{figure}

\begin{figure}
\centerline{\psfig{figure=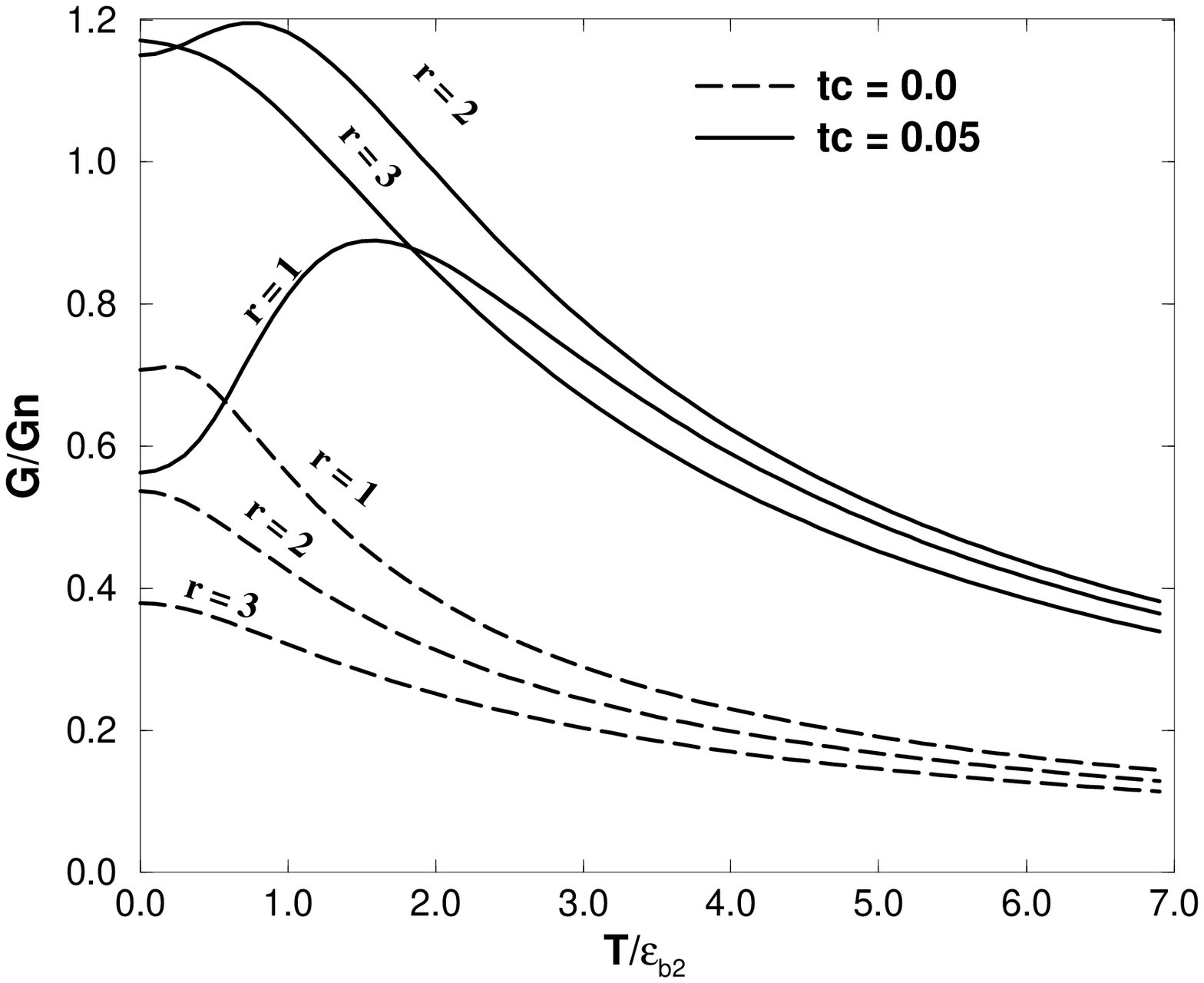,width=10cm,height=5cm}}
\caption{Temperature dependence of the zero-bias conductance of
the $N-s-S\;$structure for\\ $t_c=0.0\;$and$\;0.05,\;\epsilon _{b2}/\Delta
_S=0.05:\;r=1\;,r=2\;,r=3\;.$}
\label{Fig.4}
\end{figure}

\begin{figure}
\centerline{\psfig{figure=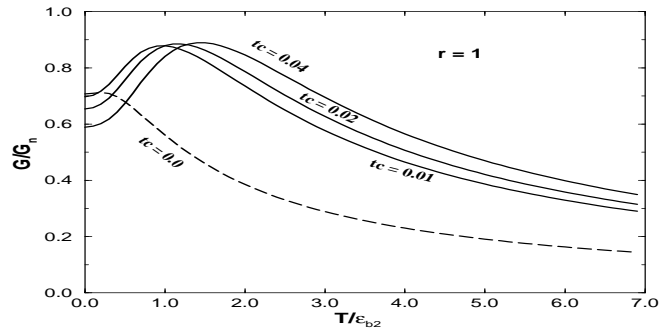,width=10cm,height=5cm}}
\caption{Temperature dependence of the zero-bias conductance of
the $N-s-S\;$structure for\\ $r=1,\epsilon _{b2}/\Delta
_S=0.05:\;t_c=0.0\;,0.01\;,0.02\;,\;0.04$.}
\label{Fig.5}
\end{figure}

\begin{figure}
\centerline{\psfig{figure=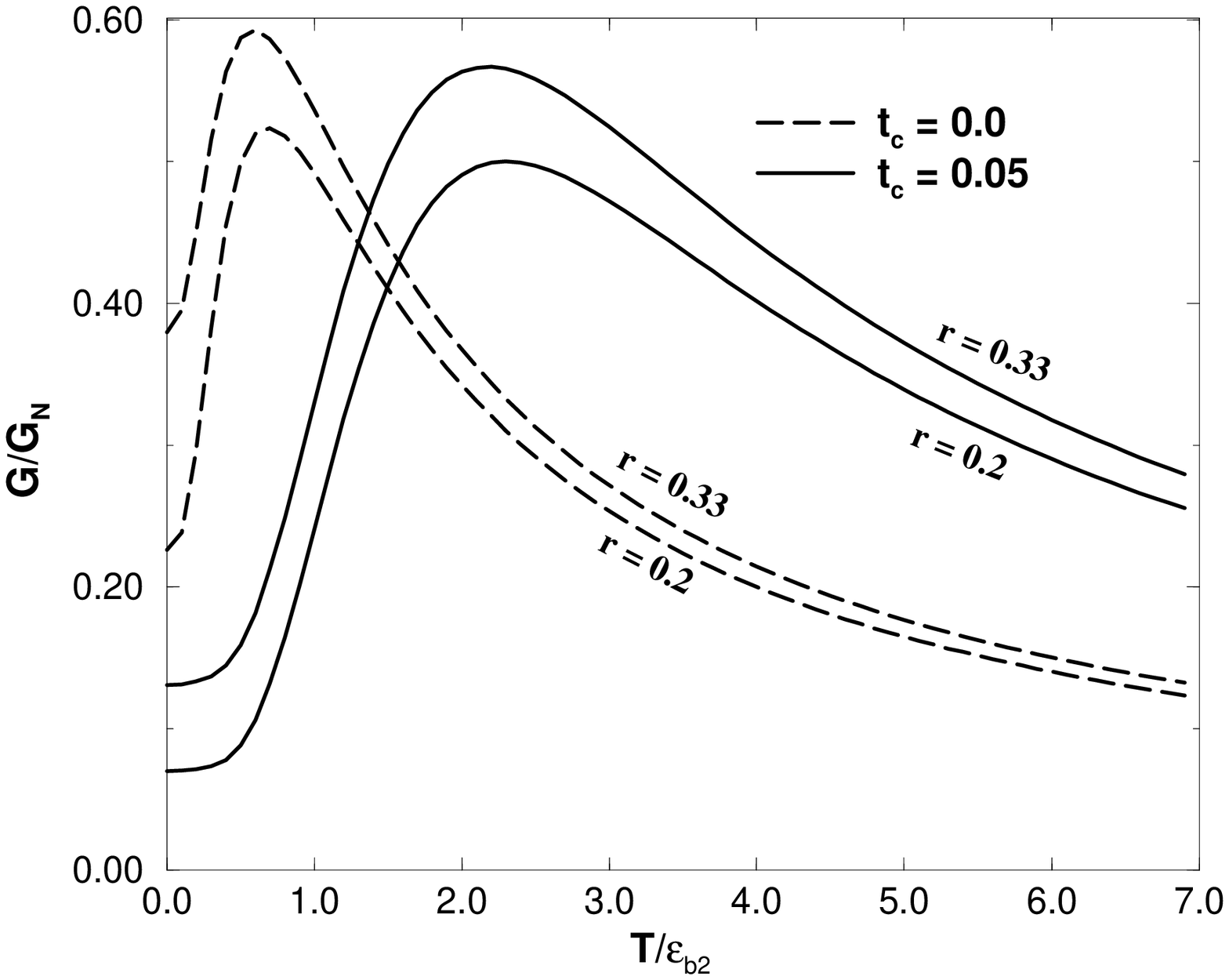,width=10cm,height=5cm}}
\caption{Temperature dependence of the zero-bias conductance of
the $N-s-S\;$structure for\\ $t_c=0.0\;$and$\;0.05,\;\epsilon _{b2}/\Delta
_S=0.05:\;r=1/5\;,\;r=1/3.$}
\label{Fig.6}
\end{figure}

\begin{figure}
\centerline{\psfig{figure=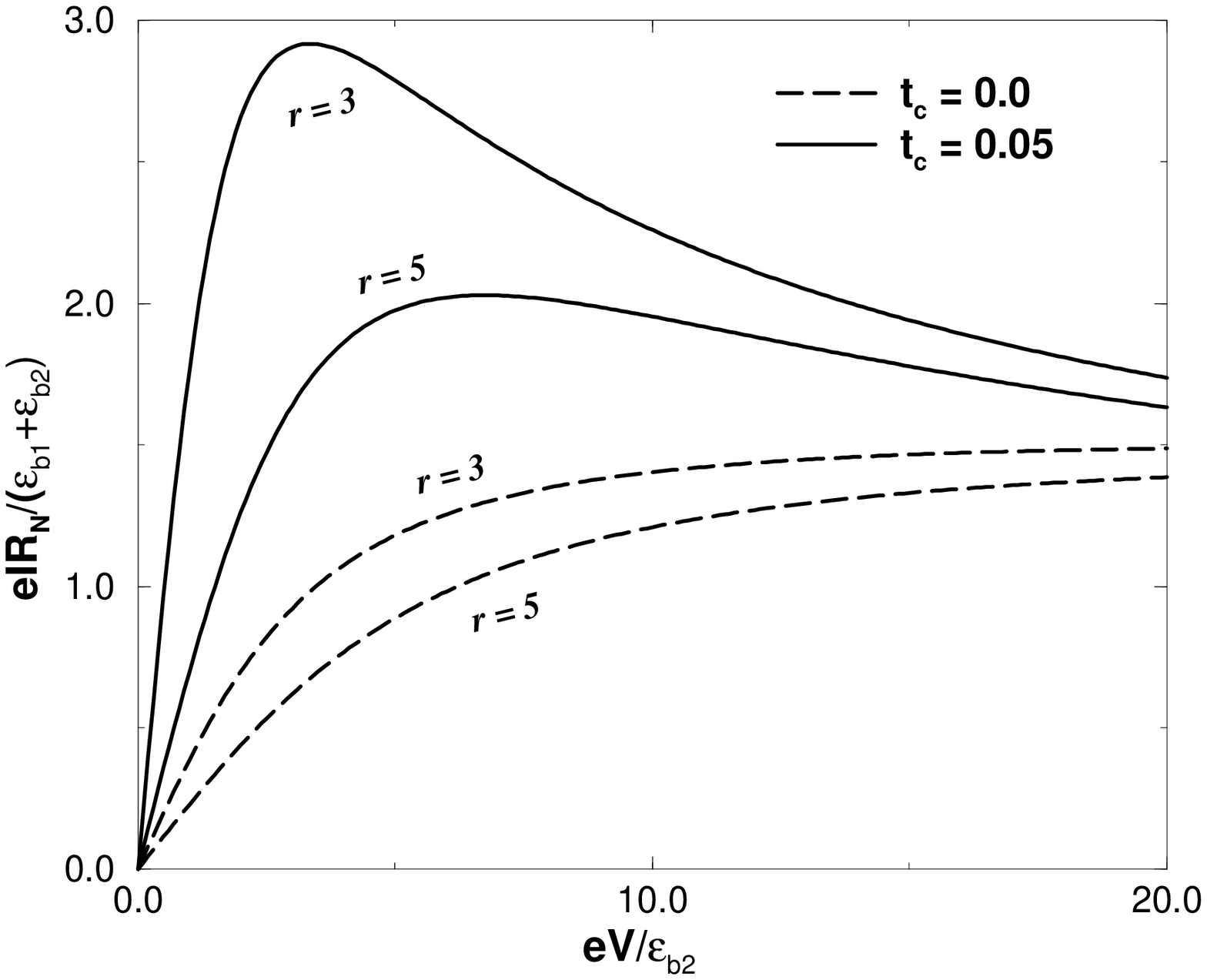,width=10cm,height=5cm}}
\caption{I-V curves at zero temperature for $r\gg1$,
$\epsilon _{b2}/\Delta_S=0.05.$}
\label{Fig.7}
\end{figure}
\end{document}